\def\year{2015}
\documentclass[letterpaper]{article}
\usepackage{aaai}
\usepackage{helvet}
\usepackage{courier}
\usepackage{balance}  % to better equalize the last page
\usepackage{graphics} % for EPS, load graphicx instead
\usepackage{times}    % comment if you want LaTeX's default font
\usepackage{url}      % llt: nicely formatted URLs
\usepackage[usenames,dvipsnames]{xcolor}
\usepackage{booktabs}
\usepackage{graphicx}
\usepackage{tabularx}
\usepackage{dcolumn}
\usepackage{enumitem}
\usepackage{subfigure}

\newcounter{JSNumberOfComments}
\stepcounter{JSNumberOfComments}

\begin{document}

\title{\#greysanatomy vs. \#yankees: Demographics and Hashtag Use on Twitter}

\author{Jisun An \quad Ingmar Weber\\
Qatar Computing Research Institute\\
Hamad Bin Khalifa University\\
\{jan, iweber\}@qf.org.qa}

\maketitle

\begin{abstract}
Demographics, in particular, gender, age, and race, are a key predictor of human behavior. Despite the significant effect that demographics plays, most scientific studies using online social media do not consider this factor, mainly due to the lack of such information. In this work, we use state-of-the-art face analysis software to infer gender, age, and race from profile images of 350K Twitter users from New York. For the period from November 1, 2014 to October 31, 2015, we study which hashtags are used by different demographic groups. Though we find considerable overlap for the most popular hashtags, there are also many group-specific hashtags. 
\end{abstract}

\section{Introduction}
Demographics are a key predictor of human behavior. The life of a 50 year old African American\footnote{Even though the official US census uses the term ``black'', we will be using ``African American'' in this paper as it is generally perceived more positively~\cite{halletal15jesp}.} woman is probably very different from that of a 16 year old white boy. The 190 billion dollar US advertising industry uses demographics to help define consumer segments that can then be targeted through dedicated campaigns\footnote{As an example, see \url{https://www.facebook.com/business/help/433385333434831} for ad targeting options provided by Facebook.}.

Despite the important effect that demographics play, most scientific studies using online social media do not consider this factor in their analysis. As an example, analyses of what is ``trending'' are generally to be interpreted as ``trending among the dominant, majority demographic group'' as the influence of the data for minorities can be too small to register. This also applies to the various ``year in review'' lists for various online services.

A key reason for this oversight is that demographic information is often not readily available and Twitter in particular lacks dedicated data fields for gender, age or race. Though a user's likely gender can often be inferred from their provided name, guessing their age and race typically involves training a classifier. For a multi-lingual corpus this can be a daunting task. In this paper we apply state-of-the-art face analysis technology to infer the demographic attributes of a user from their provided profile image. Using this approach we obtain demographic information for 346,050 Twitter users from New York.

With this large-scale data set we answer the following research question: what do different demographic groups tweet about? Are there topics that are unique to a particular group?

We hope that by showing that it is both possible and worth the effort to infer noisy-at-individual but stable-in-aggregate demographic labels for large amounts of Twitter users, more researchers will go beyond analyzing population-level behavior, as this will always be dominated by the majority groups.

\vspace{-2mm}

\section{Related Work}

Hashtags allow users to self-categorize their messages and to join a virtual conversation on a given topic. Users can search for tweets with a particular hashtag to learn about recent events on a topic of their choice. Hashtags are also frequently used in scientific studies as they are easier to obtain and handle than, say, LDA topics. A recent study on classifying hashtags and inferring semantic similarity can be found in \cite{ferragina2015analyzing}.

Many attempts have been made to understand hashtag dynamics and Lehmann et al.~(\citeyear{lehmannetal12www}) described different classes of collective attention. Lin et al.~(\citeyear{lin2013bigbirds}) characterize the growth and persistence of hashtags along four dimensions: topicality (the number of times a hashtag is retweeted), prominence (the popularity of the users mentioning the hashtag), interactivity (additional replies), and diversity (the number unique retweet sources). 
Similarly, Romero et al.~(\citeyear{romero2011differences}) studied the mechanics of information diffusion on Twitter by analyzing the spread of hashtags, focusing on the variations of the diffusion features across different topics. Cunha et al.~(\citeyear{cunha2011analyzing}) focused on linguistic characteristics of hashtags to find out what distinguishes a hashtag that spreads widely from one that fails to attract attention. 

The work closest to our study is done by Olteanu et al.~(\citeyear{olteanu2015characterizing}), who have investigated demographics of users around \#BlackLivesMatter. Though in our study we do not consider political topics separately, we also observe demographic differences for \#BlackLivesMatter, similarly to their work. Orthogonal to our work, Cunha et al.~(\citeyear{cunha2014he}) have investigated human gendered behavior in social networks, in particular attempted to verify whether the already known difference in the linguistic behavior of men and women also occur in political hashtag use. 

Concerning demographics and online behavior in general, Weber et al.~(\citeyear{webercastillo10sigir,weberjaimes11wsdm}) looked at web search, Goel et al.~(\citeyear{goeletal12icwsm}) looked at online browsing, and Malmi and Weber (\citeyear{malmiweber16icwsm}) looked a app usage.

\section{Data and Methodology}

For data collection, we first get a list of users who live in NY, then we remove those who recently joined or are not active. We then look at their profile pictures and determine their age, gender and race. For those who have a detectable face in their profile image, we collect their past tweets.

\vspace{-1mm}

\subsection{Data collection}

\textbf{Twitter Users in NY.} For our study, we focus on Twitter users who live in New York. We made this choice as (i) we wanted to limit the confounding effect of geographic-demographic changes, (ii) NY has a demographically diverse population, and (iii) there is a large number of Twitter users in NY. To gather a set of Twitter users living in NY, we use the ``search bio'' function of FollowerWonk\footnote{\url{https://moz.com/followerwonk/}}.

We used this function with the query ``$ny|nyc|brooklyn|$ $queens|yonkers|(the bronx)|(nueva york)$'' for the location field. For this query, we obtained 2,300,357 matching users in NY from FollowerWonk. For further filtering, we obtained their Twitter bios using Twitter's Restful API, resulting in 2,277,456 users. Among them, we choose ``active" users who have at least 10 tweets, joined Twitter more than 3 months ago, and have tweeted at least once in the last 3 months of the data collection period. This process leaves us with 767,300 users. 

\textbf{Demographic Inference.} For these users, we then try to infer their demographic information, concretely their age, gender, and race. There are different ways to infer gender: 1) by a gender-based dictionary, often based on census data~\cite{liu2013,amislove2011@icwsm}; 2) by a profile background~\cite{AlowibdiBY13}; 3) by tweet content, in particular for non-English languages where the form of adjectives can often reveal the gender of the speaker~\cite{cohen2013}; and 4) by web services that can be used for this purpose\footnote{\url{http://genderize.io/}}. Race can sometimes be inferred from names~\cite{amislove2011@icwsm,PennacchiottiP11,PennacchiottiP11b}. Age is harder to infer and longitudinal tweet histories would typically be required for this task~\cite{nguyen2013@icwsm}. 

We chose to use a single, language-agnostic tool that uses profile pictures to infer all three variables, age, gender and race: each profile picture, where present, was passed through the Face++ API\footnote{\url{http://www.faceplusplus.com/demo-detect/}}. When a face is detected, this API returns various bits of information, such as a gender, an age estimate, a race, whether a person is smiling or not, etc. We now discuss this process and its accuracy in more detail.

\textbf{Face detection.} Face++ detects the presence of a face for 49\% of the profile images in our data set, leaving us with 377,410 Twitter users. The detected gender for these users is fairly balanced with 51.6\% males and 48.4\% females. The majority of the users are detected as ``White'' (74.4\%), but there are also substantial fractions of ``African American'' (12.2\%) and ``Asian'' (13.4\%). We group users by their age in 6 categories: those with an inferred age below or equal to 17 (denoted as ``age-17''), 18 to 24 (age18-24), 25 to 34 (age25-34), 35 to 44 (age35-44), 45 to 54 (age45-54), and over or equal to 55 (age55-). The majority group is age25-34 which entails 37.9\% of users, then, age18-24 with 21.9\%. The smallest group is age55- with 2.3\%. The general bias towards younger adults and the very slight skew towards male users was also observed in a recent Pew Research study \cite{pewresearch2015}.

\begin{figure} [t!]
\vspace{-4mm}
  \begin{center}
  \includegraphics[width=.5\textwidth]{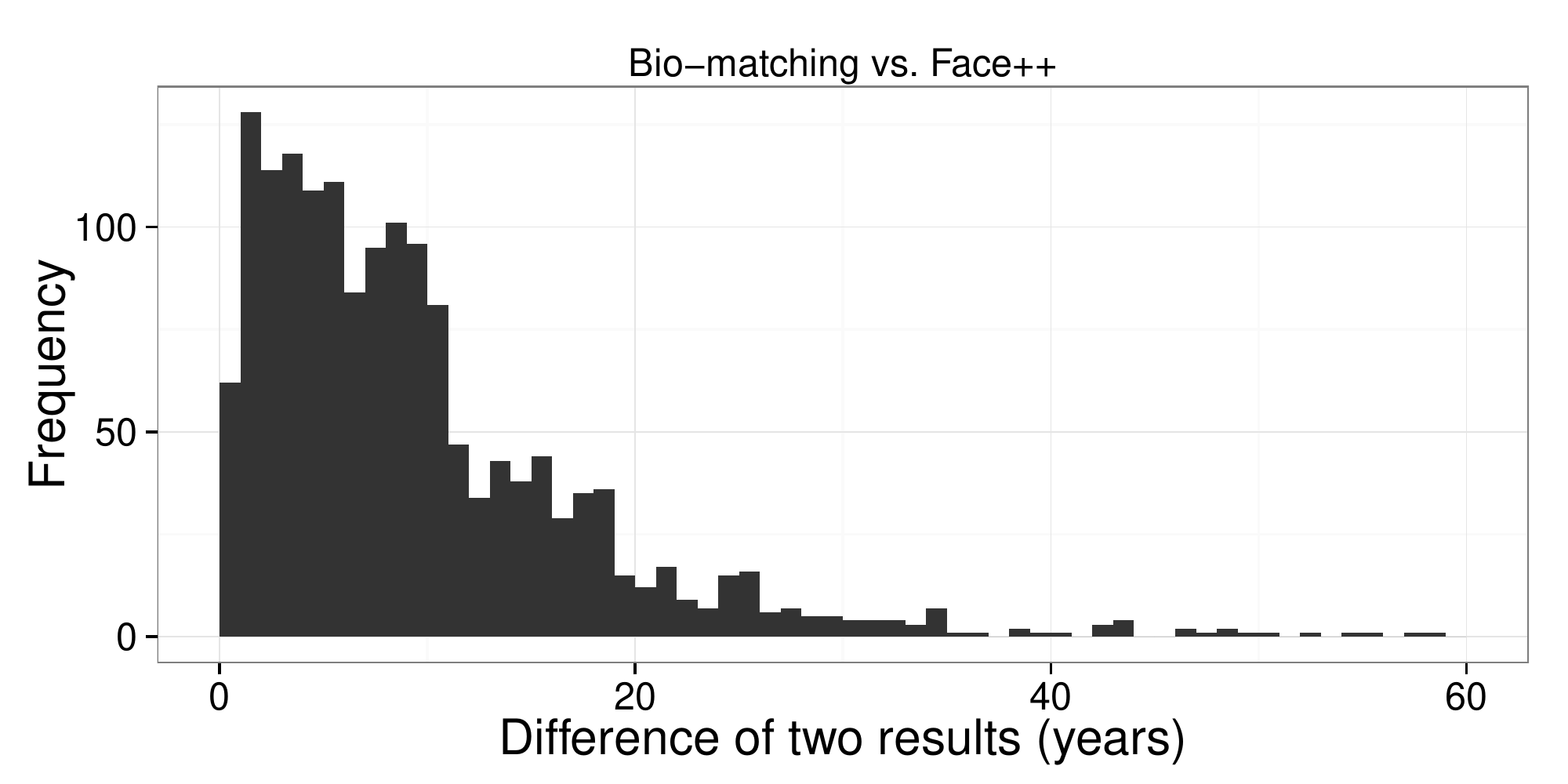}
  \caption{Histogram of the absolute age difference between the age inferred (i) by Face++ and (ii) using ``X years old'' patterns.}
  \label{fig:comparison_bio_face}
  \end{center}
  \vspace{-4mm}
\end{figure}

\textbf{Validation of Face analysis.} 
As an easy method to validate how accurate Face++ is in inferring demographic information, we look at the profile description (i.e.\ bio) of Twitter users. We tag users who describe themselves as  ``$^(boy|guy|husband|father|dad|dude)$'' as ``Male'' and $^(girl|wife|mother|mom)$ as ``Female''. For age, we look at users who explicitly mention their age (e.g., ``26 $yr|yrs|year|years$ old'').

Concerning gender, we find that of the 2,433 users with one of the female indicator terms in their bios 82\% are recognized as female by Face++. Male has an even higher detection rate--86\% of 2,033 males who use one of the male indicator terms are detected as male by Face++. 

For the age inference, the detection rate is lower than for gender since it is 6-class prediction. Figure~\ref{fig:comparison_bio_face} shows the histogram of the absolute difference between the age inferred (i) by Face++ and (ii) by regular expressions. We find that 25\% of them are within 3 year difference and 75\% of them are within 11 year difference. Although age detection through face analysis is not perfect to tell one's exact age, we believe that it is accurate enough to reason about age groups at the population level.\footnote{Note that the regular expression does also not have a perfect precision and it gets fooled by expressions such as ``11yr public school teacher'' or ``20 years experience in Macintosh support''.}

\textbf{Collecting tweets and hashtag.} For the 377,410 users for demographic information could be inferred, we collect their tweets posted during a one year period (2014/11/01--2015/10/30) using Twitter's Restful API. We collected 156,719,072 tweets for 346,050 users (92\% of users whose face was detected). 31,360 users were discarded as they no longer exist in Twitter or are protected accounts.

\subsection{Hashtags as Topic Proxies}
Hashtags are often good indicators of the topic of a tweet. Hashtags can be largely self-explanatory (e.g., \#sports) or they can be more cryptic (e.g., \#ipl standing for Indian (cricket) Premier League). In this paper, we use differences in the hashtag usage between different demographic groups as an indicator of different topical interests. Conceptually, our analysis could also be done using LDA-type topics. This would, however, involve a number of challenges, including how to avoid that the topics are dominated by the biggest demographic group. 

We extract all hashtags used in these tweets--4,648,929 hashtags used during the one year period. Of course, not all hashtags were used by many users. In fact, 73\% of hashtags (3,402,625) were used only once. The most popular hashtag is \#nyc and is used by 41,958 unique users.

\subsection{Differences in Twitter Usage}

We now look at differences in Twitter usage by marginal and joint groups. Here ``marginal'' and ``joint'' are terms from multi-variate statistics and refer to (i) only considering one demographic variable at a time, e.g.\ females, and (ii) considering a combination of demographic variables, e.g.\ 18-24 year old white males.

To quantify differences among groups for individual activity levels, we used two measures: the number of tweets per user and the tweet interval of a user. The tweet interval is approximated by the duration between the first and the last tweets divided by the number of tweets minus 1.
Individual activity levels are not strongly related to the group size. Across all users, a user posts 447 tweets during the data collection period on average (minimum is 1, maximum is 5,185 and median is 137 tweets). Asian and African American tweet more than White (the median values are 151, 164, and 131, respectively). Also younger users are likely to tweet more than elders (172, 156, 131, 112, 111, 108 from age-17 to age55-, respectively). 

Similar patterns are observed for the tweet intervals. Asian, African American and young users tend to post tweets within a much shorter interval. The median interval time is 1.7 days for Asian, 1.6 for African American, 1.6 for age below 24 and 2.3 for age over 35. Across all users, the median of access interval is 2.0 days (6.1 days as a mean).

\begin{table}[t!]
\begin{center}
\tiny \frenchspacing
\begin{tabular}{cccccc}
\toprule
\textbf{Rank} & \textbf{All} & \textbf{Female} & \textbf{African American} & \textbf{age45-54}\\
\midrule
1 & nyc & nyc & nyc & nyc\\
2 & tbt & tbt & tbt & 1\\
3 & 1 & 1 & 1 & tbt\\
4 & love & love & blacklivesmatter & 2\\
5 & 2 & lovewins & love & love\\
6 & lovewins & 2 & empire & usa\\
7 & blessed & blessed & soundcloud & periscope\\
8 & brooklyn & blacklivesmatter & ferguson & gopdebate\\
9 & blacklivesmatter & truth & np & newyork\\
10 & newyork & brooklyn & 2 & art\\
11 & ferguson & ferguson & blessed & ferguson\\
12 & music & newyork & brooklyn & brooklyn\\
13 & truth & art & music & ff\\
14 & usa & family & repost & blacklivesmatter\\
15 & art & summer & rip & lovewins\\
16 & superbowl & music & truth & music\\
17 & gopdebate & superbowl & facts & mets\\
18 & periscope & repost & wcw & superbowl\\
19 & neverforget & wcw & family & truth\\
20 & family & grammys & 2015 & travel\\

\bottomrule
\end{tabular}
\end{center}
\vspace{-4mm}
\caption{Top 20 hashtags in each group.}
\vspace{-3mm}
\label{tab:top20_hashtags_in_eachgroup}
\end{table}

\begin{table*}[th!]
\begin{center}
\tiny \frenchspacing
\begin{tabular}{p{0.4cm}p{3.2cm}p{3cm}p{2.7cm}p{2cm}p{1.8cm}p{1.8cm}}
% \begin{tabular}{ccccccc}
\toprule

\textbf{Rank} & \textbf{Female} & \textbf{Male} & \textbf{Asian} & \textbf{African American} & \textbf{White} \\
\midrule
1 & love & knicks & freshofftheboat & empire & nyr \\
2 & makeup & nfl & aldubebforlove & empirefox & sabres \\
3 & greysanatomy & jets & aldubebtamangpanahon & growingupblack & buffalo \\
4 & lovewins & mets & aldubmostawaiteddate & blacklivesmatter & bills \\
5 & beauty & nbafinals & kpop & facts & usa \\
6 & pll & nba & aldub & betawards2015 & nhl \\
7 & girlpower & soundcloud & asian & betawards & billsmafia \\
8 & nyfw & sctop10 & ootd & sandrabland & snl40 \\
9 & internationalwomensday & mlb & thepersonalnetwork & soundcloud & onebuffalo \\
10 & sisters & yankees & asianamerican & np & mets \\
\midrule
\textbf{Rank} & \textbf{age-17} & \textbf{age18-24} & \textbf{age25-34} & \textbf{age35-44} & \textbf{age45-54} & \textbf{age55-} \\
\midrule
1 & dontjudgechallenge & goals & nyc & leadership & tcot & tcot \\
2 & todayskidswillneverknow & relationshipgoals & brooklyn & innovation & iran & irandeal \\
3 & growingupwithstrictparents & todayskidswillneverknow & latergram & data & p2 & p2 \\
4 & nowimmad & mcm & nofilter & bigdata & pjnet & pjnet \\
5 & growingupagirl & wcw & broadway & iot & irandeal & ccot \\
6 & growingupwithsiblings & blessed & newyork & constantcontact & china & constantcontact \\
7 & pll & vmas & summer & mobile & leadership & uniteblue \\
8 & kca & pll & inspiration & analytics & gop & googlealerts \\
9 & ifwedate & growingupwithsiblings & sunset & jobs & constantcontact & iran \\
10 & teenchoice & tbt & tbt & digital & 8217 & ibdeditorials \\
\bottomrule
\end{tabular}
\end{center}
\vspace{-4mm}
\caption{Most discriminative hashtags for each group, ranked by Phi.}
\vspace{-3mm}
\label{tab:top10_hashtags_in_eachgroup_byphi}
\end{table*}
\textbf{Hashtags usage.} Of the 156M tweets in our dataset, 23.9\% (37M) contain one or more hashtags. When looking at marginal groups, the fraction of tweets with hashtag were similar across race (White (24.7\%), Asian (21.7\%), African American (22.1\%)) and gender groups (Female (24.4\%) and Male (23.4\%)). For age groups, the use of hashtag is increasing for older people (18.8\% for age-17 group, then 28.9\% for age45-54 group). This might hint at a more informational, rather than social use of Twitter for older users.

When looking at joint groups, the top five groups with the highest hashtag-tweet rates are: age55-+Female+Afr-American (39.5\%), age45-54+Female+White (32.1\%), age35-44+Female+White (31.5\%), age45-54+Female+ Asian (29.4\%), age55-+Female+White (29.1\%). 
We find that there is no statistically significant correlation between the fraction of tweets with hashtag and the group size--both Person's correlation and for Spearman rank correlation tests showed not significance.

\subsection{Basic observations} \label{sec:basic_observations}
\noindent \textbf{The top 20 most popular hashtags.} To gain first insights into how the hashtag usage differs between different demographic groups, Table~\ref{tab:top20_hashtags_in_eachgroup} shows the top 20 hashtags for different groups ranked by the number of unique users. Not surprisingly, the top hashtag are similar across groups with the hashtag \#nyc being the most popular.

Despite the overlap in the most popular hashtags, there are also notable differences in the rankings. For example, \#lovewins is 5th for Female but 10th for Male. 
Regarding the Ferguson shooting\footnote{\url{https://en.wikipedia.org/wiki/Shooting_of_Michael_Brown}}, \#ferguson made the top 20 list for all groups, with a ranking between 8th to 13th. \#blacklivesmatter, even though related to \#ferguson, is used differently by different groups. It was ranked highest for African American (4th) and lowest for White (19th). A related analysis was presented in \cite{olteanu2015characterizing}. 
Some hashtags only make the top 20 of particular groups. For example, \#vmas (a.k.a.,``Video Music Awards'') only appears in the age-17 top 20. For African American the hashtag \#np (9th) for ``now playing'' is unique and only age45-54 has \#travel (20th). More examples can be found in Table~\ref{tab:top20_hashtags_in_eachgroup}.

\vspace{1mm} 
\noindent \textbf{Most discriminative hashtags:} The lists of top 20 hashtags displayed mostly similarities between the groups. Here we focus on the group-specific hashtags, specifically on those with a high Chi-square score \cite{casella2002statistical} for discriminating between the group and the non-group (e.g., Female vs.\ Non-Female (Male) or Asian vs.\ Non-Asian).
Table~\ref{tab:top10_hashtags_in_eachgroup_byphi} shows the top 20 hashtags ranked by Phi, the Chi-square test statistics.
For most groups the discriminative terms are intuitive such a \#asianamerican for Asian or \#growingupblack for African American.

\section{Conclusions}
This paper is the first large-scale study presenting details on how different hashtags are used by different demographic groups on Twitter. Aggregated across a whole year the most popular hashtags are largely similar for all groups.
Our work shows that a population-level analysis of hashtags and trends on Twitter is likely to miss the complexities induced by demographic-specific behavior.

\balance

\small
\bibliographystyle{aaai}
\bibliography{icwsm2016-poster-demographics}
\end{document}